\documentclass{jpsj-suppl}
\usepackage{txfonts} 

\def\bra{\langle}
\def\ket{\rangle}

\def\Jpsi{{J\!/\!\psi}{}}
\def\X{\text{$X$(3872)}}

\def\cbar{\overline{{c}}}
\def\Dbar{\overline{{D}}{}}

\def\DDbarz{$D^0\Dbar^{*0}$}
\def\DDbarpm{$D^+D^{*-}$}
\def\ccbar{$c\cbar$}
\def\DDbar{$D\overline{D}{}^*$}

\def\rmd{{\rm d}}

\def\Belle{\text{Belle}}
\def\BABAR{\text{BaBar}}

\title{Radiative Decays of the $X(3872)$ in the Charmonium-Molecule Hybrid Picture}

\author{Sachiko \textsc{Takeuchi}$^{1}$  Makoto \textsc{Takizawa}$^{2}$ and Kiyotaka \textsc{Shimizu}$^{3}$}

\inst{$^1$Japan College of Social Work, Kiyose, Tokyo 204-8555, Japan\\
$^2$Showa Pharmaceutical University, Machida, Tokyo 194-8543, Japan\\
$^3$Department of Physics, Sophia University, Chiyoda-ku, Tokyo 102-8554, Japan}

\email{s.takeuchi@jcsw.ac.jp}

\recdate{}

\abst{The \X\ radiative decay is discussed by employing 
a two-meson multichannel model with the charmonium components,
which explains many of the other observed features of \X.
We have found that the ratio of the branching fractions of the \X\ radiative decays,
 $R_\gamma
=\text{Br}(X(3872)\!\rightarrow\!\psi(2S)\gamma)$ / Br$(X(3872)\!\rightarrow\!\Jpsi\gamma)$,
is 1.1 $\sim$ 3.4 if one assumes 
that the decay occurs only from the charmonium components.
The invariant mass spectra of $c\cbar(2P)\rightarrow \Jpsi \gamma$ and $c\cbar(2P)\rightarrow \psi(2S) \gamma$ are also investigated. It is found that
an enhancement appears at around $E\sim 3960$ MeV in the $\psi(2S) \gamma$ spectrum
but not in the $\Jpsi\gamma$ spectrum.
This corresponds to the missing $c\cbar(2P)$ pole which was predicted by a quark model 
but has not been observed due to the coupling to the $D\Dbar$ states.
We argue that 
the fluctuation seen in the experimental $\psi(2S) \gamma$ spectrum reported by LHCb
may correspond to this \ccbar($2P$) pole.}

\kword{X(3872), radiative decay, exotic hadron}

\begin{document}
\maketitle

\section{Introduction}

The \X\ peak has been found first by \Belle\ \cite{Choi:2003ue}
in the $\Jpsi \pi\pi K$ observation from the $B$ decay,
which was confirmed by
various experiments  \cite{Agashe:2014kda}.
%
The mass of \X\ is found to be 3871.69$\pm$0.17 MeV,
which is very close to or even corresponds to the \DDbarz\ threshold, 
3871.80$\pm$0.12 MeV, within the experimental errors.
The spin parity is found to be $J^{PC}$=$1^{++}$.
Some  of this \X's features cannot be understood
by a simple $q\overline{q}$ meson picture.
For example, the observed  total width is  $< 1.2$ MeV,
which is very small for an excited state.
A large isospin symmetry breaking is 
found in the decay:
the decay fraction of \X\ into $\pi^+\pi^-\Jpsi$
is comparable to that into $\pi^+\pi^-\pi^0 \Jpsi$ as
\begin{align}
{Br(X\rightarrow \pi^+\pi^-\pi^0 \Jpsi)
\over 
Br(X\rightarrow \pi^+\pi^-\Jpsi)}
&=1.0 \pm 0.4 \pm 0.3 &&(\Belle)
\label{eq:eq1}\\
&=0.8\pm 0.3 &&(\BABAR).
\nonumber
\end{align}
These two features can be nicely explained by a two-meson model with a
charmonium component; {\it i.e.},
the charmonium-molecule hybrid model \cite{Takizawa:2012hy, Takeuchi:2014rsa}.
The size of the \ccbar\ component is small, but plays an important role.
It produces a short range attraction
 for the
isospin 0 channel, 
whereas the threshold mass difference prefers the maximal isospin breaking
at the long range.
It is found that the \X\ can be a shallow bound state, which is very large in size,
or a $S$-wave virtual state.

 The absolute value of the \X\ radiative decay 
 has not been reported yet, but the relative width is reported as
\begin{align}
R_\gamma = 
{Br(X\rightarrow \psi(2S)\gamma)
\over 
Br(X\rightarrow \Jpsi\gamma)}
&= 3.4\pm 1.4 ~(3.5\sigma) ~~~&&(\BABAR[5])
\label{eq:eq2}\\
& < 2.1 ~(90\% CL)&&(\Belle[6])\nonumber\\
&=2.46\pm 0.64(\text{stat})\pm 0.29(\text{sys})  ~ (4.4\sigma) ~&&(\text{LHCb}[7]).
\nonumber
\end{align}
The data from these three experiments seem not to be inconsistent,
though  \BABAR\ and LHCb prefer a larger value whereas  Belle prefers a smaller one.
It is important to investigate the above ratio in order to understand  the nature of the \X, the most well-investigated 
exotic meson.
The theoretical works have been done with an assumption that
the \X\ is a  bound state \cite{Takizawa:2014nma,Swanson:2004pp,Barnes:2005pb, Dong:2009uf}.
In this work, we study the radiative decays of the \X\
with the above hybrid model.
Our approach here enables us to deal with the \X\ as a resonance 
and to  produce the mass spectrum of $\Jpsi\gamma$ and $\psi(2S)\gamma$.

\section{Model}
In the present model,  the \X\
 consists of \DDbarz, \DDbarpm, $\Jpsi\omega$, $\Jpsi\rho$, and 
the \ccbar(1P) ($\chi_{c1}(1P)$)  and  
 \ccbar(2P)  components.
The radiative decay can occur from each of these components.
Here we assume that the decay occurs 
only from  the \ccbar($nP$) components 
and neglect the other ones
as a first step   \cite{Takizawa:2014nma}.
The present results, especially the absolute values,  may change 
when the decay from the two-meson components are included.
We, however, consider that the characteristic feature found in 
the decay energy spectrum will probably remain unchanged
if the other decay modes are included.

The the $E1$ transition from the \ccbar\ components in the bound \X\ to the final $\psi$, which stands for $\Jpsi$ or $\psi(2S)$,
can be written as 
 \begin{align}
\Gamma \Big(X(3872)\rightarrow \psi + \gamma \Big)
&=
{4\over 9} |Q_c|^2 \alpha\;
{\omega_\gamma^3 E_\psi\over M_X}
\Big| \,
Z_{c\cbar(1P)}\bra\psi |\,r\,|{c\cbar}(1P)\ket
+
Z_{c\cbar(2P)}\bra\psi |\,r\,| {c\cbar}(2P)\ket
\,\Big|^2 \; 
%
\label{eq:eq3}
 \end{align}
where 
$Q_c$ is the electric charge of the charm quark,
$\alpha$ is the fine-structure constant,
$\omega_\gamma$ and $E_\gamma$ is the energy of the final $\gamma$ and $\psi$,
respectively,
$M_X$ is the \X\ mass,
$Z^2_{c\cbar(nP)}$ is the  probability of each \ccbar($nP$) component in the \X.

\begin{table}[bt]
\caption{The factor  $\bra \psi  |\,r\,|c\cbar(nP)\ket$ calculated by the quark model (QM) as well as the harmonic oscillator wave function with the size parameter $b$ (H.O.)  are shown.
The decay width $\Gamma$ of \ccbar($nP$) to $\psi$ 
by the quark model wave function  are also listed with 
the experimental value in parentheses. }
\label{t2}
\renewcommand{\arraystretch}{1.2}
\begin{tabular}{lccccc}
\hline
& $\bra \Jpsi|\,r\,|c\cbar(1P)\ket$ 
& $\bra \Jpsi|\,r\,|c\cbar(2P)\ket$ 
& $\bra \psi(2S)|\,r\,|c\cbar(1P)\ket$ 
& $\bra \psi(2S)|\,r\,|c\cbar(2P)\ket$ \\ \hline
QM [fm]
&  0.33 & 0.04 & $-$0.41 & 0.52
 \\
H.O.\
& $\sqrt{3\over 2}b$ 
& 0
& $-b$
& $\sqrt{5\over 2}b$
 \\
\hline
& $\Gamma(c\cbar(1P)\rightarrow \Jpsi)$
& $\Gamma(c\cbar(2P)\rightarrow \Jpsi)$
& $\Gamma(c\cbar(1P)\rightarrow \psi(2S))$
& $\Gamma(c\cbar(2P)\rightarrow \psi(2S))$
\\ \hline
QM [keV] &
207 (285) & 21 & - & 157\\\hline
\end{tabular}
\end{table}
The transfer matrix element $\bra \psi|r|{c\cbar(nP)}\ket$ 
is calculated by using the quark model wave function \cite{Godfrey:1985xj}.
Their values 
and the corresponding radiative decay widths
are listed in Table \ref{t2} together with the matrix elements evaluated by
the harmonic oscillator wave function.
The width for the $n=1$ and $\psi=\Jpsi$ case can be calibrated from the observed $\chi_{c1}(1P)$ radiative decay width, which is shown in Table \ref{t2} in  parentheses.
Note that the matrix element between the \ccbar($2P$) and $\Jpsi$  is very small;
it is zero if the harmonic oscillator wave function is employed.
This means that the radiative decay to the $\Jpsi\gamma$ mode (and also the ratio $R_\gamma$) is 
sensitive to the size of the $\chi_{c1}(1P)$ component in the \X, which is very small.
Also, in order to see the \ccbar($2P$), 
one has to look into the final $\psi(2S)\gamma$ decay mode or into the
difference between the final $\psi(2S)\gamma$ and $\Jpsi\gamma$ decay modes.

The relative phase of the  \ccbar($1P$) and the \ccbar($2P$) components in the \X\ wave function
has not been determined yet. 
We calculated three extreme cases:
there is no \ccbar($1P$) component (case A00), the decay from the two charmonia is constructive (case A01), and destructive (case A10).
All the parameter sets give correct mass the \X, 3871.69 MeV.
The mass of the \ccbar($1P$) reduces 
by 0.1 MeV for A01 and by 22.6 MeV for A10 due to the coupling to \DDbar.
The results are summarized in Table \ref{t3}.

We assume \ccbar($2P$) is created by the weak decay of the $B$-meson, 
and this \ccbar($2P$) in turn decays 
into the  final $\psi\gamma$ state: 
$B\rightarrow c\cbar(2P)K$
and then  $c\cbar(2P)$($\rightarrow $\X)
$\rightarrow \psi\gamma$.
Here we calculated the latter half of this process, $c\cbar(2P)\rightarrow \psi\gamma$,
and compare it to the experimental $\psi\gamma$ invariant mass spectrum.
The transfer strength  of this process, d$W$/d$E$,
 can be calculated as follows.
 \begin{align}
{\rmd W(c\cbar(2P)\rightarrow \psi\gamma)\over \rmd E}
&= -{1\over \pi}{\sl Im}\; \bra c\cbar(2P)| G_Q{}^{\gamma} |c\cbar(2P)\ket
\label{eq:eq4}
\\
&=\delta(E_\psi+\omega_\gamma-E)
\sum_\epsilon
\Big|\sum_n\bra \psi(E_\psi) \gamma(\omega_{\gamma},\epsilon) |
V_{\gamma Q}
|c\cbar(nP)\ket\bra c\cbar(nP)|G_{Q}|c\cbar(2P)\ket
\Big|^2
\label{eq:dWdE}
 \end{align}
where $G_Q{}^{\gamma}$[$G_Q$] is the full propagator of the \ccbar\  state
with [without] the radiative decay term,
$V_{\gamma Q}$ is the effective transfer potential from the \ccbar($nP$) to the $\psi\gamma$
state.
The parameter set we use to calculate the $\bra c\cbar(nP)|G_{Q}|c\cbar(2P)\ket$ term
is essentially the same as parameter set A in Ref.\  \cite{Takeuchi:2014rsa}.
Note that the effects of the $\rho$ and $\omega$ meson widths are included in
the self energy $G_Q$.

We evaluate the factor  $\bra \psi(E_\psi) \gamma(\omega_{\gamma},\epsilon) |
V_{\gamma Q}
|c\cbar(nP)\ket$ 
from the matrix element $\bra\Jpsi\gamma |\,r\,| {c\cbar}(1P)\ket$ by using 
the decay from the \ccbar\ without the coupling to the continuum:
\begin{align}
\Gamma \Big(c\cbar(nP)\rightarrow \psi + \gamma \Big)
&=
\sum_\epsilon
\Big|\bra \psi(E_\psi) \gamma(\omega_{\gamma},\epsilon) |
V_{\gamma Q}
|c\cbar(nP)\ket\Big|^2_{E_{\psi}+\omega_\gamma-M_{c\cbar(nP)}}
\\
&=
{4\over 9} |Q_c|^2 \alpha\;
{\omega_\gamma^3 E_\psi\over M_{c\cbar(nP)}}
\Big|\bra\psi\gamma |\,r\,| c\cbar(nP)\ket\Big|^2~.
 \end{align}

%
 
\section{Results and discussion} 
\begin{table}[tb]
\caption{The probability of the \ccbar\ components in the \X,
the radiative decay width $\Gamma$ in keV, and the ratio $R_\gamma$ are shown
for each parameter set, A00, A01, and A10.
}
\label{t3}
\renewcommand{\arraystretch}{1.2}
\begin{tabular}{lccccccccc}
\hline
model & $g_{c\cbar(1P)- {D\overline{D}}}$ & $Z^2_{c\cbar(1P)}$& $Z^2_{c\cbar(2P)}$
& $\Gamma(X\!\rightarrow\!\Jpsi)$ & $\Gamma(X\!\rightarrow\!\psi(2S))$
& $R_\gamma$ & $R_\gamma$ (spectrum) \\ \hline
A00 & 0
&0 & 0.036 & 0.6 & 2.1 & 3.6 & 3.4 \\
A01 & ${1\over 10}g_{c\cbar(2P)- {D\overline{D}}}$
&0.001 & 0.036 & 1.1 & 2.0 & 1.8 & 1.9 \\
A10 &$-g_{c\cbar(2P)- {D\overline{D}}}$
&0.011 & 0.060 & 6.1 & 6.2 & 1.0 & 1.1 \\
\hline
\end{tabular}
\end{table}
The obtained widths 
are $\Gamma(X(3872) \rightarrow \Jpsi\gamma)$ = 0.6-6.1 keV and $\Gamma(X(3872)\rightarrow\psi(2S)\gamma)$ = 2.1-6.2 keV
when we assume that the \X\ is a bound state and use eq.\ (\ref{eq:eq3}).
The ratio, $R_\gamma$, becomes 2.1-6.2, which is close to 
the \BABAR\ and LHCb results.
This, however, does not mean that our results exclude the Belle value
because the ambiguity due to the unknown relative phase of the
two \ccbar\ components is very large 
and because
the results may change
when we introduce the radiative decay from the two-meson components.

\begin{figure}[htbp]
\begin{center}
\includegraphics[width=5cm]{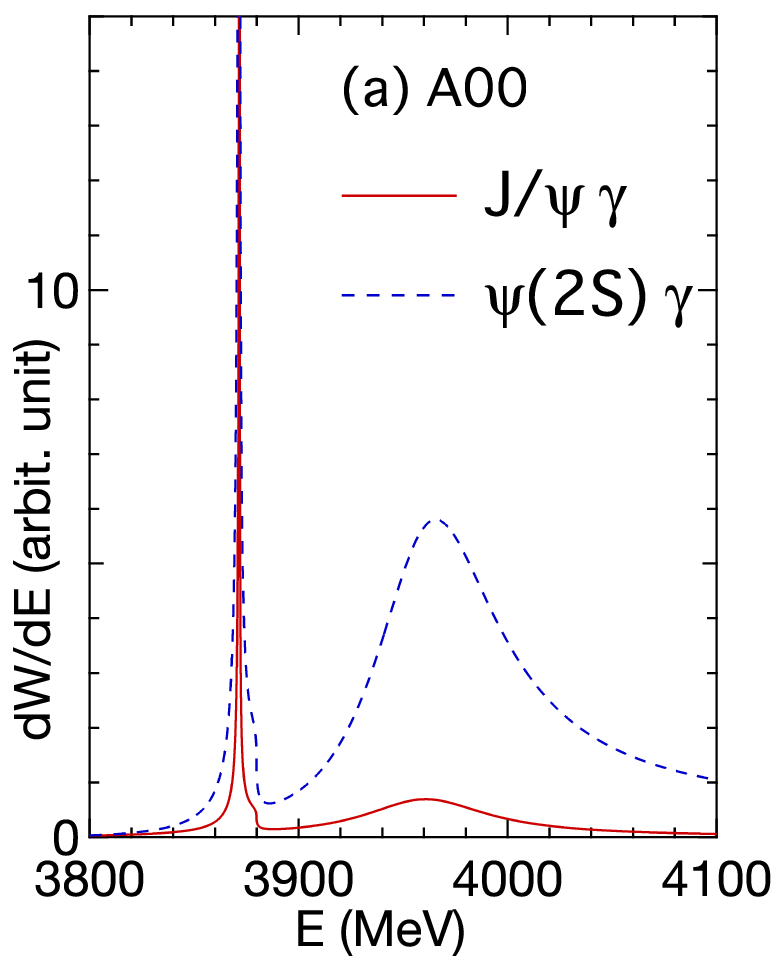}
\includegraphics[width=5cm]{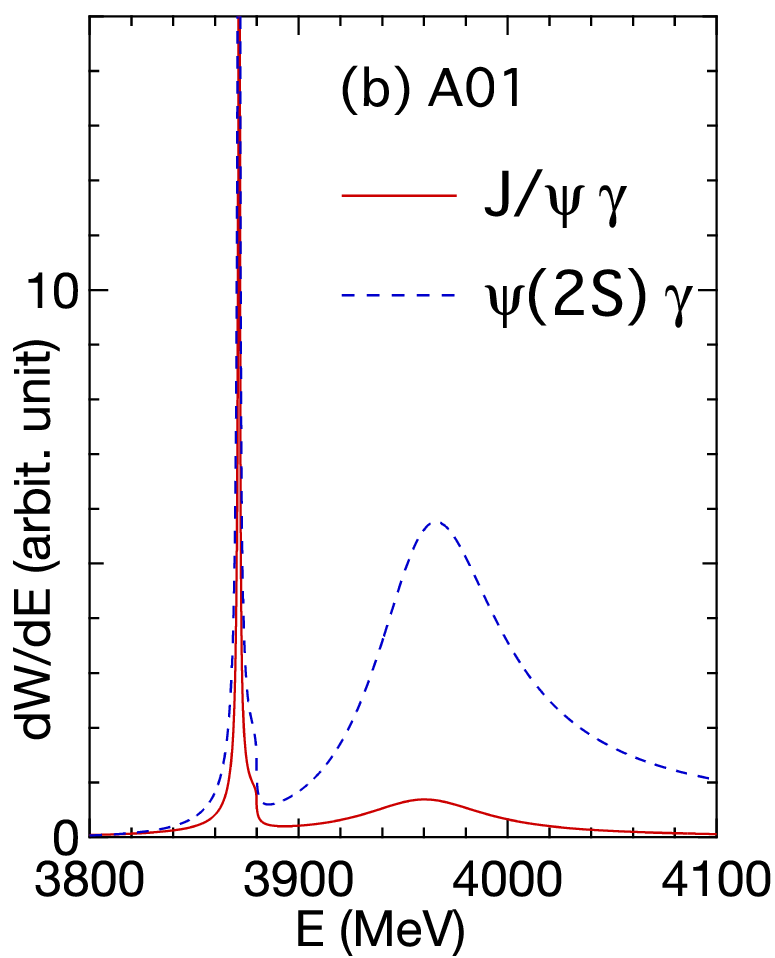}
\includegraphics[width=5cm]{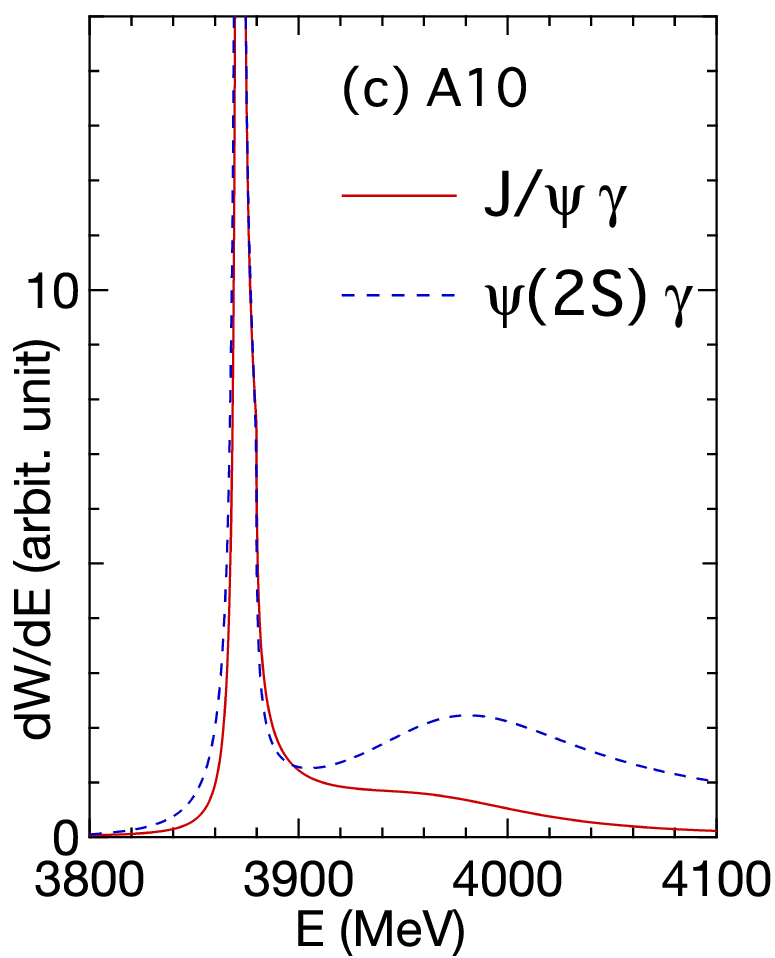}
\caption{The \ccbar($2P$) $\rightarrow \Jpsi\gamma$ and the \ccbar($2P$) $\rightarrow \psi(2S)\gamma$
mass spectra calculated by the present work are shown. Figure (a) corresponds to the parameter set A00, (b) to A01, and (c) to A10.}
\label{fig}
\end{center}
\end{figure}

The  $\Jpsi\gamma$ and the $\psi(2S)\gamma$ mass spectra
are shown in Figure \ref{fig}.
Both of the final $\Jpsi\gamma$ and $\psi(2S)\gamma$ decay have 
a peak at the energy which corresponds to the \X\ mass.
We calculated the radiative decay width of the \X\  
by integrating the strength of this peak up to the
$D^+\overline{D}{}^-$ threshold, 3879.87 MeV.
The ratio $R_\gamma$ is listed in Table \ref{t3} 
under the entry $R_\gamma$ (spectrum).
The results are not very different from those with the bound state approach.
%
In addition to the peak at the \X\ mass, 
there is an enhancement in the $\psi(2S)\gamma$ mass spectrum 
at around 3500-4000 MeV.
This enhancement occurs because of the \ccbar($2P$) pole, which
exists at $3959-{i\over 2}72$ MeV for the A00 and A01 parameter sets while it moves to
$3969-{i\over 2}140$ MeV for the A10 parameter set.
Since the \ccbar($2P$) state decays only weakly to $\Jpsi\gamma$,
as seen in Table \ref{t2},
such a structure is not seen in the $\Jpsi\gamma$ mass spectrum.
This enhancement is considered to survive
if we introduce the decay from the two-meson channels
 because the effects of the \ccbar($2P$) pole on the two-meson channels are an indirect one:
 \ccbar($2P$) $\rightarrow$ \DDbar\ $\rightarrow\psi\rho\rightarrow \psi\gamma$;
the \ccbar($2P$) contributes also to the final $\Jpsi\gamma$ mode,
but their peak structure will be much weaker.
The existence of this \ccbar($2P$) pole
 has been predicted by quark models but has not been observed because  it strongly couples 
to the
open charm channels.
We would like to point out that 
the fluctuation at around 3500-4000 MeV in the final $\psi(2S)\gamma$ mass spectrum 
(but not seen in the final $\Jpsi\gamma$ mass spectrum)
measured by LHCb \cite{Aaij:2014ala} may correspond to this enhancement.
If that is the case,
it will give us a novel method to reveal heavy quarkonia embedded in the continuum.

To evaluate the radiative decay of the \X, 
it is  necessary to look into the decay from the two-meson components
in addition to that from the \ccbar\ components.
Since the relative phase between \ccbar($1P$) and \ccbar($2P$) affects the 
results largely, this should be determined by considering a more fundamental theory.
%
Also, the consistent picture among many exotic mesons should be constructed;
looking into the radiative decay such as $Y(4260)\rightarrow X(3872)\gamma$ 
may help to understand the exotic mesons.
It should also be interesting to know how the situation changes if the charm quark
is replaced by the bottom quark.
\bigskip

\vfill

\begin{thebibliography}{99}

\bibitem{Choi:2003ue} 
  S.~K.~Choi {\it et al.}  [Belle Collaboration],
  Phys.\ Rev.\ Lett.\  {\bf 91}, 262001 (2003).

\bibitem{Agashe:2014kda}
  K.~A.~Olive {\it et al.} [Particle Data Group Collaboration],
  Chin.\ Phys.\ C {\bf 38} (2014) 090001.

\bibitem{Takizawa:2012hy} 
  M.~Takizawa and S.~Takeuchi,
  Prog. Theor. Exp. Phys.\ {\bf 2013} (2013) 0903D01.

\bibitem{Takeuchi:2014rsa} 
  S.~Takeuchi, K.~Shimizu and M.~Takizawa,
  Prog. Theor. Exp. Phys.\ {\bf 2014} (2014) 123D01.

\bibitem{Aubert:2008ae} 
  B.~Aubert {\it et al.}  [BaBar Collaboration],
  Phys.\ Rev.\ Lett.\  {\bf 102}(2009) 132001 .

\bibitem{Bhardwaj:2011dj} 
  V.~Bhardwaj {\it et al.}  [Belle Collaboration],
  Phys.\ Rev.\ Lett.\  {\bf 107}(2011) 091803 .
   %

\bibitem{Aaij:2014ala} 
  R.~Aaij {\it et al.}  [LHCb Collaboration],
  Nucl.\ Phys.\ B {\bf 886} (2014) 665.

\bibitem{Takizawa:2014nma} 
  M.~Takizawa, S.~Takeuchi and K.~Shimizu,
  Few Body Syst.\  {\bf 55} (2014) 779.
 
\bibitem{Swanson:2004pp}
  E.~S.~Swanson,
  Phys.\ Lett.\ B {\bf 598} (2004) 197.

\bibitem{Barnes:2005pb}
  T.~Barnes, S.~Godfrey and E.~S.~Swanson,
  Phys.\ Rev.\ D {\bf 72} (2005) 054026.
  
\bibitem{Dong:2009uf}
  Y.~Dong, A.~Faessler, T.~Gutsche and V.~E.~Lyubovitskij,
  J.\ Phys.\ G {\bf 38} (2011) 015001.
  
  
\bibitem{Godfrey:1985xj} 
  S.~Godfrey and N.~Isgur,
  Phys.\ Rev.\ D {\bf 32}, 189 (1985).

\end{thebibliography}
\end{document}